\documentstyle[twoside,fleqn,espcrc2,epsfig]{article}

\newcommand{\AmS}{{\protect\the\textfont2
  A\kern-.1667em\lower.5ex\hbox{M}\kern-.125emS}}

\newcommand{\nub}{\overline{\nu}}
\def\nim#1#2#3  {{\em Nucl. Instr. Meth.} {\bf#1}, #2 (#3). }
\def\np#1#2#3   {{\em Nucl. Phys.} {\bf#1}, #2 (#3). }
\def\pl#1#2#3   {{\em Phys. Lett.} {\bf#1}, #2 (#3). }
\def\prep#1#2#3 {{\em Phys. Rep.} {\bf#1}, #2 (#3). }
\def\prev#1#2#3 {{\em Phys. Rev.} {\bf#1}, #2 (#3). }
\def\prl#1#2#3  {{\em Phys. Rev. Lett.} {\bf#1}, #2 (#3). }
\def\rmp#1#2#3  {{\em Rev. Mod. Phys.} {\bf#1}, #2 (#3). }
\def\rpp#1#2#3  {{\em Rep. Prog. Phys.} {\bf#1}, #2 (#3). }
\def\zp#1#2#3   {{\em Zeit. Phys.} {\bf#1}, #2 (#3). }
\def\epj#1#2#3   {{\em Eur. Phys. Jour.} {\bf#1}, #2 (#3). }

\begin{document}

\hyphenation{author another created financial paper re-commend-ed}

\title {
A Measurement of  $xF_3^{\nu}$-$xF_3^{\nub}$ and $R$ 
        with the CCFR Detector \thanks{ To be published 
in proceedings of the 7th International
Workshop on Deep Inelastic Scattering and QCD, Zeuthen, Germany, Apr. 1999.}}

\author{U.~K.~Yang,  S.~Avvakumov, P.~de~Barbaro,  A.~Bodek,
H.~Budd,  D.~A.~Harris, K.~S.~McFarland,  W.~K.~Sakumoto\address{Department 
of Physics and Astronomy, University of Rochester, Rochester, NY 14627},
R.~A.~Johnson, M.~Vakili, V.~Wu\address{Department of Physics, 
University of Cincinnati, Cincinnati, OH 45221},
C.~G.~Arroyo, A.~O.~Bazarko, J.~M.~Conrad, J.~A.~Formaggio, 
J.~H.~Kim, B.~J.~King, S.~Koutsoliotas,
W.~C.~Lefmann, C.~McNulty, S.~R.~Mishra, A.~Romosan, 
F.~J.~Sciulli, W.~G.~Seligman, M.~H.~Shaevitz, 
P.~Spentzouris, E.~G.~Stern, B.~M.~Tamminga, A.~Vaitaitis\address{Department 
of Physics, Columbia University, New York, NY 10027 },
R.~H.~Bernstein,  L.~Bugel, M.~J.~Lamm, W.~Marsh, P.~Nienaber, 
J.~Yu\address{Fermi National Accelerator Laboratory, Batavia, IL 60510 },
T.~Adams, A.~Alton, T.~Bolton, J.~Goldman, M.~Goncharov, 
D.~Naples\address{Department of Physics, Kansas State University, 
Manhattan, KS 66506},
L.~de~Barbaro,  D.~Buchholz, H.~Schellman,  G.~P.~Zeller\address{Department 
of Physics and Astronomy, Northwestern University, Evanston, IL 60208 },
J.~Brau, R.~B.~Drucker, R.~Frey,  D.~Mason\address{Department of Physics, University of Oregon, Eugene, OR 97403 },   
T.~Kinnel, W.~H.~Smith\address{Department of Physics, 
University of Wisconsin, Madison, WI 53706}
        				 }

\begin{abstract}
We report on a measurement of the neutrino-nucleon and antineutrino-nucleon
differential cross sections in the CCFR detector.
The measurement of the differential cross sections over a wide range of
energies allows $\Delta xF_3 = xF_3^{\nu}$-$xF_3^{\nub}$ 
and $R$ to be extracted. 
$\Delta xF_3$ is related to the difference between the contributions 
of the strange and charm seas in the nucleon to production of
massive charm quark.
The results for $\Delta xF_3$ are
compared to various massive charm NLO QCD models. The $Q^2$ dependence of $R$
for $x<0.1$ has been measured for the first time. [PREPRINT FERMILAB-Conf-99/193-E,
UR-1574, ER-40685-934]
\end{abstract}

\maketitle

\section{Introduction}

Wide band neutrino beams provide an unique opportunity
to measure $R$, the ratio of the longitudinal and transverse structure 
functions, and the difference between  $xF_3^{\nu}$ and $xF_3^{\nub}$.
A measurement of $R$ provides a test of perturbative QCD at large $x$,
and probes the gluon density at small $x$.
A non-zero $\Delta xF_3=xF_3^{\nu}-xF_3^{\nub}$ originates from the
difference between the contributions of the
strange and charm seas in the nucleon to
massive charm quark production.
Thus, a measurement of $\Delta xF_3$ can be used 
to extract the strange sea, and to understand massive
 charm production
in neutrino-nucleon scattering. Previously,
information on the strange sea came from
the exclusive dimuon events channel ($\nu N \rightarrow \mu^-cX$, 
 $c$ $\rightarrow$ $\mu^+X'$)~\cite{dimu}. The dimuon analysis
requires both an understanding of charm fragmentation and 
an acceptance correction for the detection of muons from charm decays. 
In previous extractions of structure functions~\cite{SEL}, 
a leading order slow rescaling correction
for  heavy charm production was applied in order to extract 
a theoretically corrected $F_2^{\nu}$ which could be directly
compared  with $F_2^{\mu}$. 
However, because of theoretical uncertainties in the treatment 
of massive charm production~\cite{sch_dep}, 
this correction is no longer applied in this analysis.
Furthermore, the value of  $\Delta xF_3$ 
plays a crucial role in the extraction of $F_2$ at low $x$,
where there is a long-standing discrepancy~\cite{SEL}
 between CCFR and NMC $F_2$ data.
Therefore, it is important to measure the structure functions 
without any model-dependent slow rescaling corrections.

\section{The CCFR experiment}

The CCFR experiment  collected  data
using the Fermilab Tevatron Quad-Triplet wide-band  $\nu_\mu$ and $\nub_\mu$
beam. The CCFR detector~\cite{calib} consists of an
unmagnetized steel-scintillator target calorimeter
instrumented with drift chambers, followed by a toroidally 
magnetized muon spectrometer.
The hadron energy resolution is 
$\Delta E/E = 0.85/\sqrt{E}$(GeV), and the muon momentum resolution 
is $\Delta p/p = 0.11$. By measuring the hadronic energy ($E_h$), muon
momentum ($p_\mu$), and muon angle ($\theta_\mu$), we construct
three independent kinematic variables $x$, $Q^2$, and $y$.
The relative flux
at different energies obtained from the events 
with low hadron energy ($E_h < 20$ GeV) is normalized so that
the neutrino total cross section equals the world average $\sigma^{\nu N}/E=
(0.677\pm0.014)\times10^{-38}$ cm$^2$/GeV and $\sigma^{\overline{\nu} N}
/\sigma^{\nu N}=0.499\pm0.005$~\cite{SEL}. 
The total data sample used for the extraction
of structure functions consists of
 1,030,000 $\nu_{\mu}$ and 179,000 $\nub_{\mu}$
events after fiducial and kinematic cuts ($E_{\mu}>15$ GeV, 
$\theta_{\mu} <0.150$, $E_h > 10$ GeV, and 30 GeV $<E_{\nu}<$ 360 GeV). 
Dimuon events are removed because of the ambiguous identification 
of the leading muon for high-$y$ events.

\section{Measurement of differential cross sections}

\begin{figure}[ht]
\centerline{\psfig{figure=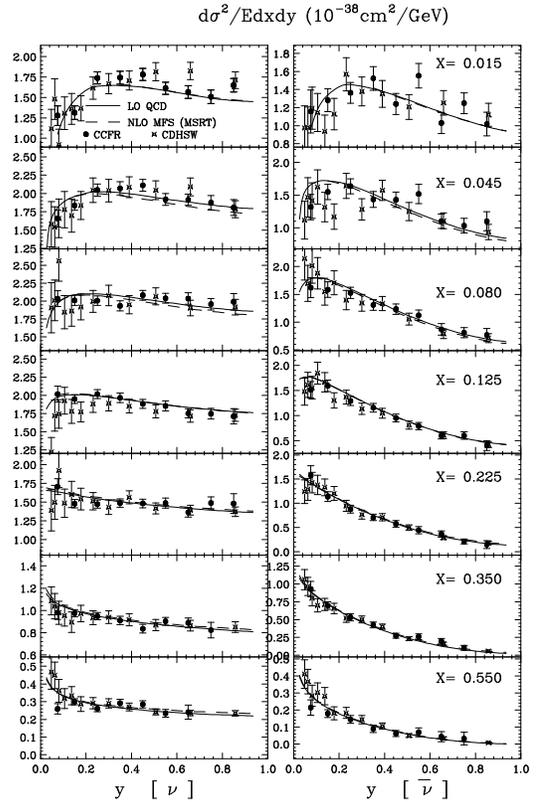
,width=2.7in}}
\vspace{-0.2in}
\caption{Preliminary CCFR differential cross section data at $E_\nu=150$ GeV
(both statistical and major systematic errors are included for the 
CCFR and CDHSW data). There is
good agreement with the NLO MFS QCD calculation using MRST PDFs,
and with LO QCD model which is used in the CCFR Monte Carlo.
A disagreement
is observed in the $y$ distribution between CCFR data and CDHSW data.}
\label{fig:diff}
\vspace{-0.15in}
\end{figure}

The differential cross sections are determined in bins of $x$, $y$, 
and $E_{\nu}$ ($0.01 < x < 0.65$, $0.05<y<0.95$, and $30< E_\nu <360$ GeV). 
Over the entire $x$ region, differential cross sections are in 
good agreement with NLO Mixed Flavour Scheme (MFS)~\cite{mfs} 
QCD calculation using the MRST~\cite{mrst} PDFs.
This calculation includes an improved treatment of massive charm
production.
Figure \ref{fig:diff} shows typical differential cross sections 
at $E_\nu=150$ GeV.
Also shown are
the prediction from a LO QCD model, which is only
used in the calculation of acceptance and resolution smearing correction.
The CCFR data exhibit a quadratic $y$ dependence at small $x$ 
for neutrino and anti-neutrino, and a flat $y$ distribution at high $x$ 
for the neutrino cross section.
Note that the $y$ distributions of the CDHSW~\cite{cdhsw} differential cross 
sections data do not agree with those of CCFR data,
or with the MRST predictions (this difference  
is crucial in any QCD analysis).

\section{Measurement of  $\Delta xF_3$ and $R$}

Values of $\Delta xF_3$ and $R$ are extracted from the sums of neutrino
and anti-neutrino differential cross sections.
The sum of the two differential cross sections can be written as:
\begin{tabbing}
$F(\epsilon)$ \= $\equiv \left[\frac{d^2\sigma^{\nu }}{dxdy}+
\frac{d^2\sigma^{\overline \nu}}{dxdy} \right]  \frac {(1-\epsilon)\pi}{y^2G_F^2ME}$ \\
  \> $ = 2xF_1 [ 1+\epsilon R ] - \frac {y(1-y/2)}{1+(1-y)^2} \Delta xF_3 $
\end{tabbing}
where $\epsilon\simeq2(1-y)/(1+(1-y)^2)$ is the polarization of virtual
$W$ boson.  To fit $\Delta xF_3$, $R$, and $2xF_1$ at a given $x$ and $Q^2$
is very challenging because of the  strong correlation 
between  the $\Delta xF_3$ and $R$ terms, unless the full range of
$\epsilon$ is covered by the data. Covering this range 
(especially the low $\epsilon$ region correspoing to high $y$) is difficult
because of the low acceptance at high $y$.

Since the contribution of $\Delta xF_3$ to $F(\epsilon)$ increases
with $Q^2$ while that from $R$ decreases,
our strategy is 
to fit $\Delta xF_3$ and $2xF_1$ for $Q^2>5$ where the $\Delta xF_3$
contribution is relatively larger,
constraining $R$ with $R_{world}$~\cite{rworld}.
($\Delta xF_3$ fits are done only for $x<0.1$, 
because its contribution is small for $x>0.1$.)
It is reasonable to use $R_{world}$
(an empirical fit for all available $R$ data),
because above $Q^2=5$, $R$ for
 neutrino and muon scattering are expected
to be the same, and $R_{world}$ is in good agreement
with the NMC muon data~\cite{nmcR}. 

In the $Q^2<5$ region, where the contribution from $R$ is larger,
we fit $R$ and $2xF_1$ with $\Delta xF_3$ constrained to the
model predictions.
This allows us to 
 study the $Q^2$ dependence of $R$ for  neutrino scattering 
at low $Q^2$. In the $Q^2>5$ region,
we test which scheme is favored for the massive charm treatment, based on
the extracted $\Delta xF_3$ results. 
Then the favored scheme is used in constraining
$\Delta xF_3$ for the $Q^2<5$ analysis.

Before the structure function extraction, we apply 
an electroweak radiative correction (Bardin) and corrections
for the $W$ boson propagator,
and a non-isoscalar target (the 6.85\% excess 
of neutrons over protons in iron; this effect is valid only at high $x$).
 The values of $\Delta xF_3$ and $R$ are sensitive to the energy
dependence of the neutrino flux, but are insensitive to the
 absolute normalization.
The uncertainty on the flux shape is estimated by using the 
constraint that $F_2$
and $xF_3$ should be flat over $y$ (or $E_{\nu}$) for each $x$ and $Q^2$ bin.

\begin{figure}[ht]
\centerline{\psfig{figure=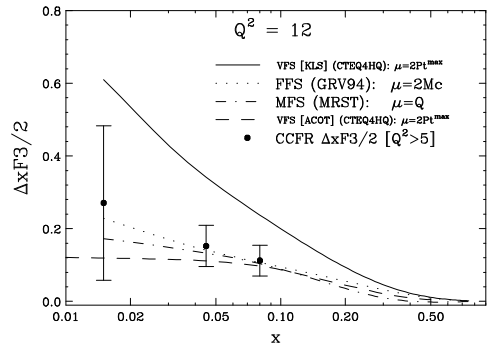,width=2.5in}}
\vspace{-0.2in}
\caption{Preliminary CCFR $\Delta xF_3$ data as a function of $x$ 
compared with various schemes for massive charm production: 
KLS's VFS with CTEQ4HQ, FFS with GRV94, 
MFS with MRST, and ACOT's VFS with CTEQ4HQ}
\label{fig:dxf3}
\vspace{-0.15in}
\end{figure}

In the $\Delta xF_3$ fit, we use the NLO
MFS calculation with MRST PDFs for the $Q^2$ dependence of $\Delta xF_3$
(for $Q^2 > 5$), and fit for the level of $\Delta xF_3$
for each $x$ bin below $x=0.1$.
Figure~\ref{fig:dxf3} shows the $\Delta xF_3/2$ data as a function of $x$.
The CCFR data favor the Fixed Flavour Scheme (FFS)~\cite{ffs} calculation with
the GRV94~\cite{grv94} PDFs,  and also the Mixed
Flavour Scheme (MFS) using MRST PDFs. 
The data do not favor the VFS calculation with CTEQ4HQ~\cite{cteq4hq}
as implemented
 by Kramer, Lampe, and Spiesberger (KLS)~\cite{sch_dep}. 
The difference among the various schemes does not come from the different
parton distributions, but rather from the different treatments of massive charm. 
In fact, all strange sea distributions in these parton distributions
agree with the CCFR dimuon data~\cite{dimu}.
Recently Olness and Kretzer~\cite{private} pointed out that
 KLS did not include the LO  charm sea diagram
in the VFS implementation, as required in the original ACOT's VFS
calculation~\cite{vfs}.
The ACOT's VFS calculations agree with data
after the inclusion of the  charm sea contribution, as
shown in Fig.~\ref{fig:dxf3}.

\begin{figure}[ht]
\centerline{\psfig{figure=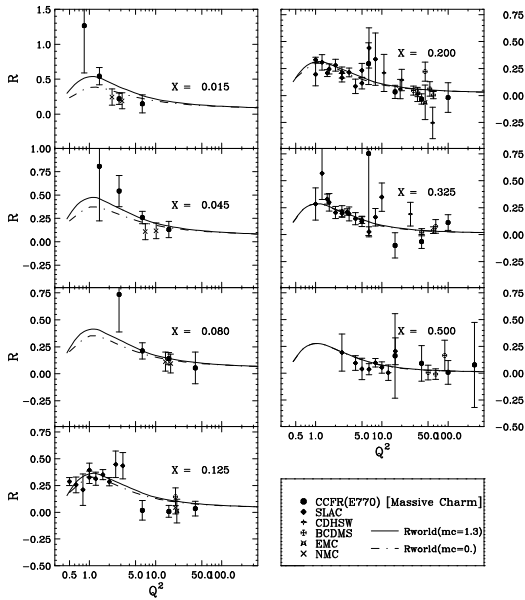,width=2.7in}}
\vspace{-0.2in}
\caption{Preliminary CCFR $R$ data as a function of $Q^2$ for various $x$, 
compared with other data and $R_{world}$ with $m_c=0$ and $m_c=1.3$.}
\label{fig:R}
\vspace{-0.15in}
\end{figure}

Since all three schemes except KLS'
VFS show a good agreement with data, 
we use a MFS calculation with  MRST PDFs
(arguably the most reasonable theoretical 
description of massive charm production currently available) 
to constrain 
$\Delta xF_3$ for the extraction of $R$ for $Q^2<5$ and $x<0.1$.
Even for $Q^2>5$, we extract $R$ in order to investigate
its $Q^2$ dependence.
The extracted values of $R$ at fixed $x$ vs $Q^2$ are shown 
in Fig.~\ref{fig:R}. 
The new data reveal the $Q^2$ dependence of $R$ at $x<0.1$ 
for the first time ($R$ decreases as $Q^2$ increases for fixed $x$). 
The NMC data shown in Fig.~\ref{fig:R} are integrated over $Q^2$,
and the two nearest NMC $x$ bins are shown together. 
At higher $x$, our measurements 
agree with the other world data for $R$~\cite{rworld,nmcR,slacr,otherr}. 
Figure~\ref{fig:R} also shows
a comparison with $R_{world}$ with $m_c=0$ (muon scattering 
or with slow rescaling correction in neutrino scattering) and $m_c=1.3$
(without slow rescaling correction in neutrino scattering).
The CCFR $R$ data at $x=0.015$ do not show $R$ approaching zero
as $Q^2$ goes zero (as expected in neutrino scattering, but not in
electron scattering).

\section{Conclusions}
New measurements of $\Delta xF_3$ and $R$ have been
extracted from the CCFR differential cross section data. The $R$ data
are extended to lower $x$ and higher $Q^2$ than previous measurements.
The $Q^2$ dependence of $R$ at lower $x$ region 
has been measured for the first time.
The $\Delta xF_3$ data from $Q^2>5$ region agree with 
various schemes for the treatment of massive charm production. 
Further reduction of the
 errors in $\Delta xF_3$ is expected by including lower $Q^2$ data.
 The effect of $\Delta xF_3$ on the extraction of $F_2$ is
currently under study.
In addition, new data from the recent NuTeV run (1996-97), taken with 
sign selected neutrino beams, are expected to yield more precise
determinations of $\Delta xF_3$, $F_2$ and $R$ at low $x$.

\end{document}